\documentclass[twocolumn,floatfix,superscriptaddress,amsmath,showpacs,showkeys,aps,prb]{revtex4-1}
\usepackage[utf8]{inputenc}
\usepackage[final]{graphicx}
\usepackage{bm}
\usepackage[normalem]{ulem}
\usepackage[usenames]{color}
\usepackage{times}
\usepackage{verbatim}
\usepackage{xcolor}
\usepackage{hyperref}
\usepackage{wrapfig}
\usepackage{subfigure}
\DeclareGraphicsExtensions{.eps}

\begin{document}
\title{Two-dimensional Fermi liquid dynamics  with density and  quadrupolar interactions}
\author{Rui Aquino}
\affiliation{Departamento de F{\'\i}sica Te\'orica,
Universidade do Estado do Rio de Janeiro, Rua S\~ao Francisco Xavier 524, 20550-013  
Rio de Janeiro, Brazil}
\author{Daniel G. Barci}
\affiliation{Departamento de F{\'\i}sica Te\'orica,
Universidade do Estado do Rio de Janeiro, Rua S\~ao Francisco Xavier 524, 20550-013  
Rio de Janeiro, Brazil}
\date{\today}

\begin{abstract}
We consider a Fermi liquid model with density-density as well as quadrupolar forward scattering  interactions parametrized by the Landau parameters 
$F_0$ and $F_2$. Using bosonization and 
a decimation technique, we compute collective modes and spectral functions for a huge range of interactions, ranging from strong repulsion to 
strong attraction in either angular momentum channels.  We present a dynamical phase diagram showing a region of parameters where the collective 
modes structure changes abruptly, possibly signaling a dynamical phase transition.  
\end{abstract}

\maketitle

\section{Introduction}
\label{Sec:Introduction}
Quadrupolar  interactions in Fermi liquids are increasingly been considered  due to the possible relevance  in 
understanding the strange metal behavior observed in several highly correlated electron systems~\cite{FradKiv2010}. 
Attractive quadrupolar interactions trigger an electronic isotropic-nematic quantum phase transition~\cite{OgKiFr2001}.
This transition could be at the bottom of  anisotropic transport properties observed in several materials, running from cuprates 
superconductors~\cite{Daou2010}, 
pnictides~\cite{Tanatar2016}, quantum Hall devices~\cite{FrKi1999} and  heavy fermions systems~\cite{Borzi2007}.  

Even though a systematic classification of the isotropic-nematic quantum critical point is still lacking, the order parameter dynamics 
is reasonably well understood. In particular,  the most relevant contribution to the Fermionic dynamics at criticality seems to  come from 
an overdamped collective excitation with cubic dispersion relation~\cite{Lawler2006}.
Moreover, superconducting instabilities in the presence of quadrupolar interactions open the possibility of very interesting states of 
matter~\cite{BeFrKi2009,BarciFradkin2011, BaClSi2016} with novel topological excitations combining fractional vortex and disclinations.

In the usual theory of Fermi liquids~\cite{nozieres-1999}, two-body forward scattering interactions are parametrized by  Landau parameters 
$F_\ell$, with $\ell=0,1,2,\ldots$, where the index $\ell$ labels angular momentum representations. For instance,  while $F_0$ parametrizes 
the isotropic density-density interaction, $F_2$ measures the intensity of quadrupolar interactions. 
In Ref.~\onlinecite{Lawler2006}, a simple model with pure quadrupolar coupling was studied near the isotropic-nematic  phase transition 
driven by the Pomeranchuk instability, $F_2\sim -1$. Interestingly, it was observed that the Goldstone mode (in the nematic phase) has 
a precursor in the disordered isotropic phase which dominates the dynamics near the phase transition.

With this motivation in mind, we present in this paper a detailed analysis of the  collective modes structure of a more general model, taking 
into account density as well as quadrupolar interactions. 
We explore collective modes in a huge range of parameters, running from  strong repulsive  to strong attractive  interactions in both angular momentum channels,  up to the Pomeranchuk instability region.

By means of a two-dimensional Bosonization 
tech\-nique~\cite{CastroNeto-1993,CastroNeto-1994,CastroNeto-1995, houghton-1993,houghton-1994,houghton-2000,haldane-1994,BaOx2003}, 
we parametrize  Fermi surface deformations in terms of a set of chiral Bosons. We write an evolution equation in the semiclassical approximation, 
analog to the Landau Fermi liquid formalism.  Using an angular momentum basis, the model is reduced to a set of infinitely coupled harmonic oscillators. 
Each oscillator describes a Fermi surface deformation mode with a specific symmetry. In this way, the  model is mapped into  a classical chain of 
harmonic oscillators, in which the angular momentum label, $\ell$, plays the role of  lattice sites in a one-dimensional chain.   
Taking advantage of this  structure, we are able to exactly compute  Green functions using a recursive decimation  procedure~\cite{CaPa2006}; 
a particular implementation of  the real space renormalization group theory~\cite{Koiller1981,Sokoloff1982,Kadanoff1983}.
By looking for Green functions singularities, we compute  collective modes and spectral functions.  In this way, we are able to build a 
dynamical phase diagram (Fig.~\ref{fig:pd}) in the  $(F_0,F_2)$ plane. We show that, in specific regions of  this  plane, the Fermi surface 
dynamics changes abruptly, signaling  a dynamical phase transition.    

The main results of the paper are displayed in Figs.~\ref{fig:SF} and~\ref{fig:SFA}, in which we show longitudinal as well as transverse 
quadrupolar spectral functions for different regions of the dynamical phase diagram (Fig.~\ref{fig:pd}). In Fig.~\ref{fig:SFa}, we show the 
equivalent of the Landau zero sound for the quadrupolar interaction, expected to occur for repulsive couplings, while in the 
second panel, 
Fig.~\ref{fig:SFb}, an unexpected result is displayed. 
Due to the quadrupolar attraction, besides the Landau zero sound, another  well-defined collective mode appears at high frequencies, clearly separated from the particle-hole continuum. 
 As the attraction becomes stronger, 
these two energy levels approximate each other and finally melt  in a single damped mode at a critical value of the Landau parameters, in which the Green function has a second-order pole. This is an example of a so called ``exceptional point'' in open quantum systems~\cite{Rotter1998,Rotter2009}. At these points, eigenfunctions properties, such as orthogonality and phase rigidity suddenly change and have been associated with a dynamical phase transition (DPT)~\cite{Rotter2015}.
This sudden change in the dynamics occurs in an extended region of the parameter space (white region in Fig.~\ref{fig:pd}).
For  stronger attractive interactions, near the Pomeranchuk instability, the overdamped mode, which is present in  the whole  attractive region, 
acquires a huge spectral weight, as shown in Fig.~\ref{fig:SFc}. This mode is identified as the precursor of the nematic Goldstone mode, computed  in Ref.~\onlinecite{Lawler2006}.  Finally, Fig.~\ref{fig:SFA} displays the spectral function of transverse modes. This type of modes cannot 
exist in a Fermi liquid with 
isotropic interactions. However, they are possible when higher angular momentum components of the interactions are considered. 
Very recently, an equivalent excitation called share sound  was computed in a Fermi liquid model with  dipolar interactions ($F_1$)~\cite{Khoo2019}. 
For attractive interactions, the share mode is damped, acquiring an important spectral weight near zero frequency, in the Pomeranchuk instability region.
In the rest of the paper we give details of the model and calculations that conduce to these results. 

The paper is organized as follows: in Sec.~ \ref{Sec:FermiSurface}, we briefly review the Bosonization approach to Fermi liquids  while in 
Sec.~ \ref{Sec:Interactions} we explicitly present the model . In Sec.~\ref{Sec:Green}, we  compute the Green functions and  
in Sec.~\ref{Sec:CollectiveModes} we analyze the  structure of the collective modes. In Sec.~\ref{sec:SpectralFuncions} we show the spectral 
functions in different regimes and  we finally discuss our results in Sec.~\ref{Sec:Discussion}. We have left to the Appendix~\ref{App:Decimation} a detailed  description of the decimation technique to compute Green functions. 

\section{Fermi surface dynamics}
\label{Sec:FermiSurface}

In this section, we briefly review the bosonization approach to Fermi liquids, in order to  establish  notation, the model and the  main approximations 
we use in the paper.
 
The Fermi surface deformation can be parametrized in terms of a  set of $N$ chiral Bosons $\varphi_S({\bf x},t)$, $S=1,\ldots,N$, where the 
index $S$ labels a patch of width $\Lambda$ and height $\lambda$ in which the Fermi surface was coarse-grained. All through the paper, 
we use bold characters to represent vector quantities.    
The location of the Fermi momentum is then written as 
\begin{equation}
{\bf k}_S({\bf x},t)={\bf k}^0_S+{\bf \nabla}\varphi_S({\bf x},t)
\end{equation}  
where  ${\bf k}^0_S$ is the original uniform Fermi momentum at the patch $S$.
This parametrization should be understood in a semiclassical scheme, since it mixes momentum and configuration space in a coarse-grained 
scale  $|{\bf x}| \gg 1/|{\bf k}_F|$.
To recover the original Fermi surface, it is necessary to perform the continuum limit $\Lambda\to 0$, $N\to\infty$  with $\Lambda N=2\pi k_F$.  

A generating functional can be constructed in a  coherent state path integral formalism as\cite{CastroNeto-1995,houghton-2000} 
\begin{equation}
Z=\int {\cal D}\varphi \; e^{-i S[\varphi_S]} 
\label{eq:Z}
\end{equation}
where the integration measure is  ${\cal D}\varphi=\prod_S   {\cal D}\varphi_S $.
The action can be split into two terms,  $S=S_0+S_{\rm int}$. The first term codifies the  free Fermionic dynamics and is given by
\begin{equation}
  S_0 = \frac{N(0)}{2}\sum_S\!\int d^2xdt\bigg[-\partial_t\varphi_S{\bf v}_S
  \cdot\nabla\varphi_S-\big({\bf v}_S\cdot\nabla\varphi_S\big)^2\bigg] ,
 \label{eq:S0}
\end{equation}
  where ${\bf v}_S$ is the Fermi velocity in each patch $S$, pointing perpendicular to the surface. $N(0)$ is the density of states at the Fermi surface.  
The second one represents two-body forward scattering interactions and is given by
\begin{align}
\label{eq:Sint}
  S_{\rm int} =  \frac{N(0)}{2} \sum_{S,T}&\int d^2xd^2x'dt\;
  F_{S,T}({\bf x-x'}) \times \nonumber \\
  & \left( {\bf v}_S\cdot\nabla\varphi_S({\bf x})\right) \left({\bf v}_T\cdot\nabla\varphi_T({\bf x'})\right) \; .
\end{align} 
 $F_{S,T}({\bf x}-{\bf x}')$ represents particle-hole pair interactions between patches $S$ and $T$.
Since this is a quadratic theory, in principle, it has exact solution. Note, however, that the quadratic character has been obtained by 
linearizing the Fermionic dispersion relation around the Fermi surface. Curvature terms give rise to non-quadratic Bosonic 
interactions~\cite{BaOx2003}. Even so these terms are important to stabilize Pomeranchuk instabilities, they are irrelevant
in the Fermi liquid phase~\cite{Lawler2006}.

A semiclassical description is obtained by computing the Euler-Lagrange equations $
\delta S/\delta\varphi_S=0$, with $S=1,\ldots,N$. They can be cast in terms of  normal fluctuations
\begin{equation}
\delta n_S({\bf x},t) = N(0)\left({\bf
v}_S\cdot\nabla\varphi_S({\bf x},t)\right), 
\end{equation} 
obtaining
\begin{align}
&\frac{\partial \delta n_S({\bf q},t)}{\partial t}+ \left({\bf v}_S\cdot {\bf q} \right)
\sum_T \left\{\delta_{S,T}+
F_{S,T}({\bf q}) \right\}\delta n_T({\bf q},t)=0.
\label{eq:Boltzman}
\end{align} 
The diagonal term in the patch basis (the  $\delta$-function in the second term) 
comes from the free Fermion dynamics, while  interactions mixes patches all around the Fermi surface.

For simplicity,  we will focus on a circular Fermi surface. In this case, and in the absence of external magnetic fields, 
the angular momentum basis is much more convenient than the patch basis. Using the fact that $\delta n_S$ and $F_{S,T}$ are 
periodic functions of $S$, we can write, 
\begin{align}
\delta n_S({\bf q})&=\sum_{\ell=-\infty}^\infty m_\ell({\bf q}) \;e^{i \ell \theta_S} 
\label{eq:deltans-Fourier}  \\
F_{S,T}({\bf q})&=\sum_{\ell=-\infty}^\infty F_\ell({\bf q}) \; e^{i \ell \left(\theta_S-\theta_T\right)}
\label{eq:FST-Fourier}
\end{align} 
where $\cos(\theta_S)={\hat n}_S\cdot {\bf q}$ and $\cos(\theta_T)={\hat n}_T\cdot {\bf q}$.  Thus, $\theta_S$ ($\theta_T$) is the angle subtended 
by the  particle-hole momentum ${\bf q}$ with  the normal direction to the patch ${\hat n}_S$ (${\hat n}_T$). 
Since $F_{S,T}\equiv F(|\theta_S-\theta_T|)$, then $F_{\ell}=F_{-\ell}$.
In terms of $m_\ell({\bf q})$, the equation of motion reads, 
\begin{align}
&\frac{\partial m_\ell({\bf q},t)}{\partial t}+i\frac{v_F q}{2}\left[\alpha_{\ell-1}m_{\ell-1}({\bf q},t)+\alpha_{\ell+1}m_{\ell+1}({\bf q},t)\right]=0
\label{eq:1order}
\end{align}
in which $\alpha_\ell({\bf q})\equiv1+F_\ell({\bf q})$.

It is convenient to define symmetric and antisymmetric variables,
\begin{equation}
m^\pm_{\ell}({\bf q},t)=\frac{1}{2}\left[m_{
\ell}({\bf q}, t)\pm m_{-\ell}({\bf q},t)\right]
\end{equation}
in such a way that  $m^+_{-\ell}=m^+_{\ell}$ and $m^-_{-\ell}=-m^-_{\ell}$. 
In terms of these modes,  Fermi surface deformations are parametrized as
\begin{align}
&\delta n_S({\bf q},t)= m_0({\bf q},t)\ +\nonumber \\
&\sum_{\ell=1}^\infty \left\{ m^+_{\ell}({\bf q},t)
\cos(\ell\theta_S)+ m^-_{\ell}({\bf q},t) \sin(\ell\theta_S)\right\}.
\end{align}
Eliminating in equation (\ref{eq:1order}) odd components in favor of  even ones,
we obtain the set of coupled  differential  equations~\cite{Lawler2006,CastroNeto2005}
\begin{align}
&\frac{\partial^{2} m^\pm_{\ell}({\bf q},t)}{\partial t^{2}}+\left(\frac{v_F
q}{2}\right)^{2} \left[A_\ell\; m^\pm_{\ell}({\bf q},t)\ + \right.
 \nonumber \\
& \; \; \; + \left. C_{\ell-1}\;m^\pm_{\ell-2}({\bf q},t)+C_{\ell+1}\;m^\pm_{\ell+2}
({\bf q},t)\right]=0
\label{eq:osc}
\end{align}
with  dimensionless coefficients
\begin{equation}
A_{\ell}=\alpha_\ell(\alpha_{\ell-1}+\alpha_{\ell+1})\mbox{~~,~~}
C_{\ell}=\alpha_{\ell}\sqrt{\alpha_{\ell+1}\alpha_{\ell-1}}.
\end{equation}
The first line of  Eq.~(\ref{eq:osc}), represents an harmonic oscillation for the mode $m^{\pm}_\ell$ with angular frequency  
$\omega^2=(v_Fq/2)^2 A_\ell$. The second line couples  the angular momentum mode $\ell$ with the corresponding modes  $\ell+1$ and $\ell-1$. 
In this way, the model is mapped into a chain of harmonic oscillators with  ``first near neighbors'' couplings.  
Due to parity symmetry, symmetric and antisymmetric modes $(m^+_\ell, m^-_\ell)$, as well as odd and even angular momentum  modes, 
are completely decoupled. This structure enormously simplifies the analysis of the dynamics.  The presence of a magnetic field 
breaks parity, mixing all modes in a nontrivial way~\cite{BaRe2013,Son2016,BaFrRi2018}.

\section{Model for density and quadrupolar interactions}
\label{Sec:Interactions}
The Landau theory of a Fermi liquids can be considered as a fixed point in the renormalization group sence. In particular, it was shown~\cite{CastroNeto-1995, Shankar1994,Metzner-1998} that forward scattering interactions labeled by Landau parameters, $F_\ell$, with $\ell=0,1,2,3,\ldots$ are marginal. This result was criticized~\cite{Chitov1998}  based on the fact  that  small angle singularities could  generate, in the renormalization group flux, higher Landau harmonics, invalidating, in some sense,  simple models with  few Landau parameters. Some aspects of this idea were recently applied to  dipolar Fermi gases~\cite{Keles2016}. The  effect  of small angle singularities could be specially dangerous near Pomeranchuk instabilities, however it seems to be irrelevant in the Fermi liquid phase at reasonable weak coupling.
With this caveat in mind, and since we are interested in studying quadrupolar interactions in the stable Fermi liquid phase, we choose a Fermi liquid model whose 
interactions are parametrized by Landau parameters $F_0$ and $F_2$.  Thus, we ignore any  other component of the interaction, 
$F_\ell=0$ for $\ell\neq 0,2$. 
Then, the system of Eq.~(\ref{eq:osc}) is characterized by two  parameters $\alpha_0=1+F_0$ and  $\alpha_2=1+F_2$, 
with $\alpha_\ell=1$ for all $\ell\neq 0,2$. 

Since odd and even modes are decoupled, let us focus on even angular momentum modes, 
$m^{\pm}_0({\bf q}, t)$, $m^{\pm}_2({\bf q}, t)$, $m^{\pm}_4({\bf q}, t),\ldots$.  By arranging these modes in a vector 
form ${\bf m}^{\pm}=(m_0^{\pm}, m_2^{\pm},\ldots)$, we can rewrite Eq.~(\ref{eq:osc}), in matrix notation. Fourier transforming in time 
and defining the dimensionless variable $s=\omega/v_F q$, where $\omega$ is the frequency, we have 
\begin{equation}
\left(s^2 {\bf I} -{\bf  M}^{\pm}\right) {\bf m}^{\pm}=0,
\label{eq:I-M}
\end{equation}
where ${\bf I}$ is the identity matrix and  ${\bf M}^{\pm}$ are two independent  matrices driving the dynamics of the symmetric 
modes (${\bf M}^+$) and the antisymmetric ones (${\bf M}^-$).

We want to compute the Green functions 
\begin{equation}
{\bf G}^{\pm}(\omega,{\bf q})=\left(s^2 {\bf I} -{\bf  M}^{\pm}\right)^{-1}.
\end{equation}
Explicitly, we have 
\begin{equation}
\label{eq:M+}
\left[{\bf G}^{+}\right]^{-1} = 
\left(
\begin{array}{cc|ccc} 
    s^{2} - \frac{\alpha_0}{2} & - \frac{\sqrt{\alpha_0\alpha_2} }{2} & 0 & 0 & ... \\  
    - \frac{ \sqrt{\alpha_0\alpha_2}}{4} & s^{2} - \frac{\alpha_2}{2} & - \frac{\sqrt{\alpha_2}}{4} & 0 & \\ 
\hline     
    0 & - \frac{\sqrt{\alpha_2}}{4} & s^{2} - \frac{1}{2} & - \frac{1}{4} & \ldots\\
    0 & 0 & - \frac{1}{4} & s^{2} - \frac{1}{2} &  \\
    \vdots &  & \vdots &  & \ddots
\end{array}
\right)
\end{equation}
for the symmetric modes, and 
\begin{equation}
\label{eq:M-}
\left[{\bf G}^{-}\right]^{-1} = 
\left(
\begin{array}{c|ccc}  
     s^{2} - \frac{\alpha_2}{2} & - \frac{\sqrt{\alpha_2}}{4} & 0 & \ldots\\
     \hline
     - \frac{\sqrt{\alpha_2}}{4} & s^{2} - \frac{1}{2} & - \frac{1}{4} & \ldots\\
     0 & - \frac{1}{4} & s^{2} - \frac{1}{2} &  \\
    \vdots   &  \vdots &  & \ddots
\end{array}
\right)
\end{equation}
for the antisymmetric ones. Notice that in the antisymmetric channel there is no isotropic mode $m_0^-\equiv 0$. For this reason, 
while the first matrix acts on the space spanned by $(m_0, m_2^+, m_4^+, \ldots)$,  the second  matrix acts on the space spanned 
by $(m_2^-, m_4^-, \ldots)$.  For the same reason, the  coupling constant $\alpha_0$ does not affect the dynamics of the antisymmetric modes. 
We have explicitly indicated the block structure of Eqs.~(\ref{eq:M+}) and~(\ref{eq:M-}) by means of auxiliary vertical and horizontal lines 
inside the matrices. The first quadrant of  the matrix $[{\bf G}^+]^{-1}$ is a $2\times 2$ block, containing essentially the couplings of the modes 
$m_0$ and $m_2^+$. The fourth quadrant, in principle a block with infinite components, has no coupling constant and it is a tridiagonal matrix, 
representing  the remaining  ``free''  higher angular momentum modes. In this sense, the higher angular momentum modes are acting as a heat bath 
for the isotropic and quadrupole modes.  Both blocks are coupled just by one element which is out of diagonal blocks, by means  
of the quadrupolar interaction $\alpha_2$.  Eq.~(\ref{eq:M-}) has a similar structure, indeed a simpler one, since there is no isotropic 
coupling $\alpha_0$.

\section{Green functions}
\label{Sec:Green}
Let us focus on the computation of some elements of the Green functions ${\bf G}^\pm(s,{\bf q})$. 
We are particularly interested in the first  $2\times 2$ block  
\begin{equation}
{\bf \bar G}^\pm(\omega, {\bf q})=\left(
\begin{array}{cc}
G_{00}^{+}(\omega, {\bf q}) & G_{02}^{+}(\omega, {\bf q})  \\
G_{20}^{+}(\omega, {\bf q}) & G_{22}^\pm(\omega, {\bf q})
\end{array}
\right).
\label{eq:Gbar}
\end{equation} 
 Notice that $G_{00}^{-}=G_{02}^{-}=G_{20}^{-}=0$, since there are no antisymmetric isotropic mode  $m_0^{-}\equiv 0$.  Thus, the Green function 
 for the antisymmetric mode $m_2^{-}$ is the single function $G_{22}^{-}$. 
With these Green functions at hand,  it is possible to study time evolution of the modes $m^\pm_2(t)$ and $m_0(t)$, provided we 
establish  initial conditions in some of these two channels. 

To compute ${\bf G}^{\pm}$, we use a recursive decimation procedure 
which is an application of the real space renormalization group theory to a one-dimensional chain\cite{CaPa2006,Koiller1981,Sokoloff1982,Kadanoff1983}.  
The method is conceptually simple. We begin by truncating the matrices of Eqs.~(\ref{eq:M+}) and~(\ref{eq:M-}) to a finite range $n$.  We invert 
the finite range matrix finding a recursive relation of the type
\begin{equation}
{\bf G}^{\pm (n+1)}={\cal  F}\left[{\bf G}^{\pm (n)}\right]
\end{equation} 
where ${\bf G}^{\pm (n)}$ is the inverse of Eqs.~(\ref{eq:M+}) and~(\ref{eq:M-}), respectively, truncated to order $n$. ${\cal  F}$ is 
some matrix function.  The exact solution is found by taking the limit ${\bf \bar G}^\pm=\lim_{n\to\infty} {\bf G}^{\pm(n)}$.
In Appendix \ref{App:Decimation}, we show the explicit computation in detail. 
We find the following results,  
\begin{equation}
{\bf \bar G}^{+}(s, q)= \frac{1}{D^{+}(s,q)}\; {\bf N}^{+}(s,q),
\label{eq:G+}
\end{equation}
where the numerator 
\begin{equation}
{\bf N}^{+}(s,q)=\left(
\begin{array}{cccc}
s^2-\alpha_2(\frac{1}{2}+\Pi(s)) & & &  \frac{\sqrt{\alpha_0\alpha_2}}{2}    \\
& & & \\
 \frac{\sqrt{\alpha_0\alpha_2}}{4}    &   & &  s^2-\frac{\alpha_0}{2}\\
\end{array}
\right)
\label{eq:N+}
\end{equation}
and the denominator
\begin{equation}
D^{+}(s,q)=\left[ s^{2} - \alpha_{2}\left( \frac{1}{2} + \Pi(s) \right) \right]\left(s^{2} - \frac{\alpha_{0}}{2}\right) - 
\frac{\alpha_{2}\alpha_{0}}{8}.
\label{eq:D+}
\end{equation}
For the antisymmetric modes, we find similar results 
\begin{equation}
\bar G^{-}(s,q)= \frac{s^2-\frac{1}{2}-\Pi(s)}{D^{-}(s,q)},
\label{eq:G-}
\end{equation} 
with
\begin{equation}
    D^{-}(s,q) = \left[ s^{2} - \frac{1}{2} - \Pi(s)  \right]\left(s^{2} - \frac{\alpha_{2}}{2}\right) - \frac{\alpha_{2}}{16} \ .
\label{eq:D-}
\end{equation}
In Eqs.~(\ref{eq:N+}), (\ref{eq:D+}),  (\ref{eq:G-}) and~(\ref{eq:D-}) 
\begin{equation}
\Pi(s) = \left\{
\begin{array}{ccc}
\frac{1}{2}\left\{ s^{2} - \frac{1}{2} \pm |s|\sqrt{s^{2} - 1} \right\}  & \mbox{for} & s>1 \\
\\
\frac{1}{2}\left\{ s^{2} - \frac{1}{2} \pm i |s|\sqrt{ 1-s^{2}} \right\}  & \mbox{for} & s<1 
\end{array}
\right.
\label{eq:Pi}
\end{equation}
Please, see Appendix \ref{App:Decimation} for its derivation. 

The physical interpretation of this result is very interesting. We are computing the Green functions for the modes $m_0, m_2^{\pm}$, 
taking into account their interactions with a ``heat bath'' composed by  all other angular momentum modes.  
It is worth to note that the Green functions truncated to order $n$ (see Appendix~\ref{App:Decimation})  have  $2n$ real poles 
in the region $-1<s<1$.  In the limit $n\to \infty$, the poles are dense, giving rise to a cut for $s^2<1$, as can be seen in Eq.~(\ref{eq:Pi}). 
Moreover, for $s^2>1$, $\Pi(s)$ is real, while for $s^2<1$, it gets an imaginary part. This is nothing but Landau damping.  In next sections we 
explicitly compute collective modes and spectral functions.    

\section{Collective Modes}
\label{Sec:CollectiveModes}
In usual Fermi liquids with isotropic interactions, there are only symmetric collective modes. However, when higher Landau parameters are considered, 
antisymmetric modes  are also possible. In the following, we analyze these two cases separately. 

\subsection{Symmetric collective modes}
In order to compute symmetric collective modes, we  look for solutions of  $D^+(s)=0$, where  $D^+(s)$ is given by Eq.~(\ref{eq:D+}) 
with $\alpha_0>0$ and $\alpha_2>0$.  $\Pi(s)$ is given by Eq.~(\ref{eq:Pi}).

The first thing to note is that $D^+(0)=0$ independently of the interactions $\alpha_0$ or $\alpha_2$. This zero mode is associated with  static 
local reparametrization of the Fermi surface and, thus, it is not physical.  In this section, we will seek solutions  with $s\neq 0$. Later, 
when computing spectral functions, we  will analyze this ``spurious'' zero mode in detail. To check the consistency of the method, 
let us look for  solutions in the simplest case of $\alpha_2=1$, {\em i.e.}, a system with only density interactions $\alpha_0=1+F_0$. 
We easily  find the single solution 
 \begin{equation}
 s^2=\frac{\alpha_0^2}{2\alpha_0-1}= \frac{(1+F_0)^2}{1+2F_0} ,
 \end{equation} 
 which is the well-known Landau zero sound in two dimensions~\cite{CastroNeto2005}.

\begin{figure}[hbt]
\begin{center}
\subfigure[]
{\label{fig:roots-alpha=0-a}
\includegraphics[width=0.42\textwidth]{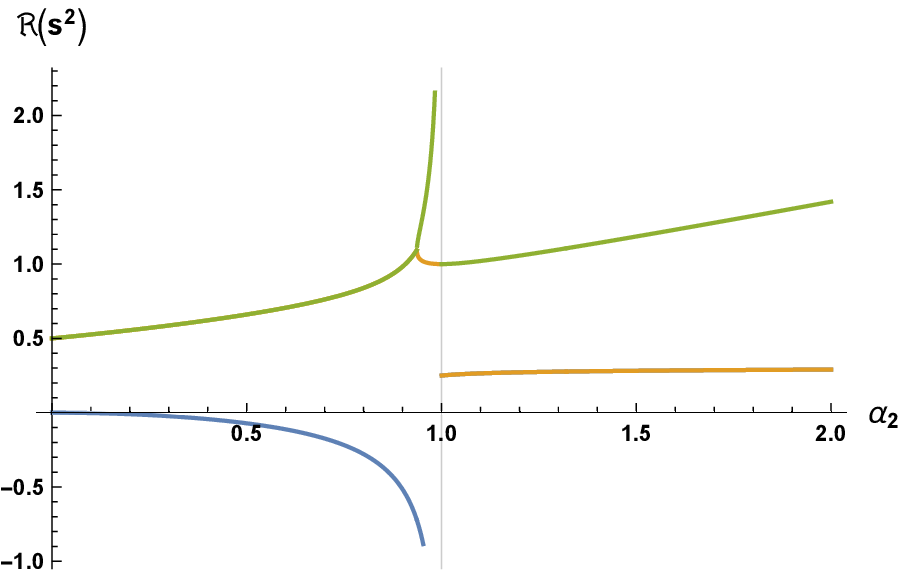}}
\subfigure[]
{\label{fig:roots-alpha=0-b}
\includegraphics[width=0.42\textwidth]{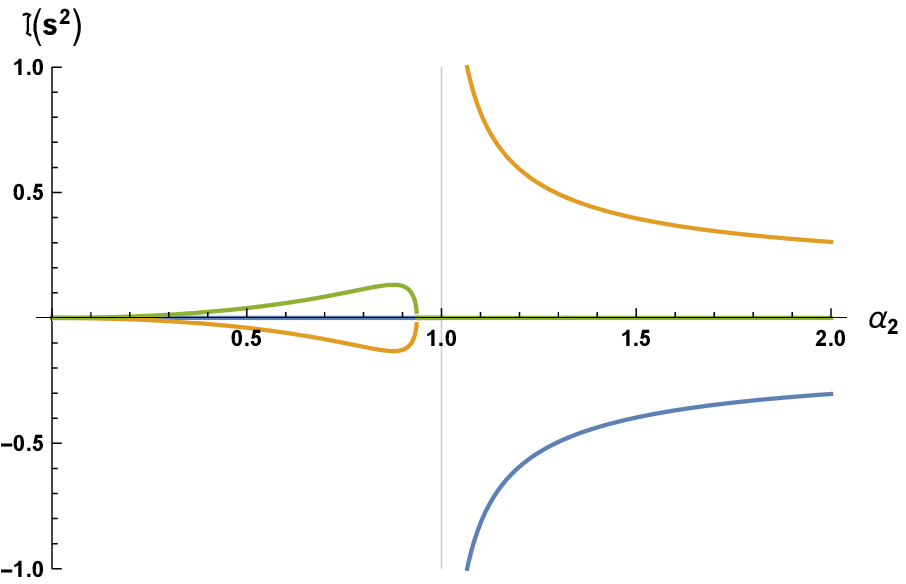}}
\end{center}
\caption{Symmetric collective modes. Solutions of the equation $D^+(s)=0$, with $D^+(s)$ given by Eq.~(\ref{eq:D+}). 
In panel~\ref{fig:roots-alpha=0-a} we plot $\Re[s^2(\alpha_2)]$ while in panel~\ref{fig:roots-alpha=0-b} we depict $\Im[s^2(\alpha_2)]$. 
In both panels we have fixed $\alpha_0=1$.}
\label{fig:roots-alpha2}
\end{figure}

Equivalent collective modes in the case of pure quadrupolar interactions, {\em i.e.}, $\alpha_0=1$ and $\alpha_2\neq 1$, can be found. 
In this case,  $D^+(s)=0$ reduces to a  cubic polynomial equation in $s^2$. 
We  depicted the roots in Fig.~\ref{fig:roots-alpha2} as a function of the quadrupolar interaction $\alpha_2$.

 In panel~\ref{fig:roots-alpha=0-a}, we show the real part of the roots, while in panel~\ref{fig:roots-alpha=0-b} we depict the 
 corresponding imaginary part. We can observe an interesting  structure. For repulsive interactions, $\alpha_2>1$, the system has one 
 real positive  root with $s^2>1$.  This is the analog of the Landau zero sound for the quadrupolar interaction.
We can give explicit analytic expressions for weak as well as strong repulsion, 
\begin{equation}
s^2=\left\{
\begin{array}{lcl}
1 + 4 (\alpha_2 - 1)^2 +\ldots & \mbox{~for~} & \alpha_2 \gtrsim 1 \\
\frac{3}{8}+\frac{1}{2}\alpha_2+O(1/\alpha_2) & \mbox{~for~} & \alpha_2 \gg 1
\end{array}
\right.
\end{equation}
Moreover, we have one additional complex root (and its complex conjugate) representing damped modes,
\begin{equation}
s^2\sim\left\{
\begin{array}{lcl}
 \frac{1}{4}\pm i\frac{1}{4 \sqrt{\alpha_2 - 1}}& \mbox{~for~} & \alpha_2 \gtrsim 1 \\
\frac{5}{16} \pm  \frac{4}{25}  i +O(1/\alpha_2)
 & \mbox{~for~} & \alpha_2 \gg 1
\end{array}
\right.
\end{equation}
This mode is inside the particle-hole continuum of the spectrum.  Very near the Fermi gas, when $\alpha_2\to 1^+$,  the real root 
$s^2\to 1$ and the imaginary part of the complex one  diverges, since in the free case only the mode 
$s^2=1$ should survive.

The attractive quadrupolar case is more interesting. In this case, there is an overdamped mode in all the attractive regime ($0<\alpha_2<1$).  
We have, 
\begin{equation}
s^2=\left\{
\begin{array}{lcl} 
-\frac{1}{4}\frac{1}{\sqrt{1-\alpha_2}} \; \hspace{0.6cm} \mbox{~for~}  \alpha_2\lesssim 1\\
\\
-\frac{\alpha_2^2}{4}   \hspace{1.5cm} \mbox{~for~}  \alpha_2\gtrsim 0
\end{array}
\right.
\label{eq:overdamped}
\end{equation}
In addition, for weak attraction, $\alpha_2\lesssim 1$, there are two stable collective modes, well separated from the particle-hole continuum. 
They take the form, 
\begin{align}
s_1^2&\sim 1+\frac{19}{4} \left(1-\alpha_2\right) + O\left(\left(1-\alpha_2\right)^2\right), \\
s_2^2&\sim \frac{1}{8}\; \frac{1}{\sqrt{1-\alpha_2}} \; + \; O\left(\left(1-\alpha_2\right)^{-3/2}\right).
\end{align}
$s_1$  is  the  continuation of the quadrupolar zero sound observed in the repulsive case to the weak attractive regime. 
On the other hand, $s_2$ is proper of attractive interactions and it has no counterpart in the repulsive regime.  
For moderate attractions  both stable modes meet each other, merging in a damped mode that remains in the strong attractive limit.
Thus, we clearly see two points where the dynamical regime changes abruptly. One point is $\alpha_2=1$, the transition between the 
repulsive and attractive regime. The other point, $\alpha_{2c}\sim 0.93$, is when  two real poles  transform in a double pole acquiring 
an imaginary part for $\alpha_2<\alpha_{2c}$.  To the best of our knowledge,  this is  an unexpected feature and is an example of an ``exceptional 
point'', studied in the literature of non-hermitian effective Hamiltonians~\cite{Rotter2015}. 

\begin{figure}[hbt]
\begin{center}
\includegraphics[width=0.35\textwidth]{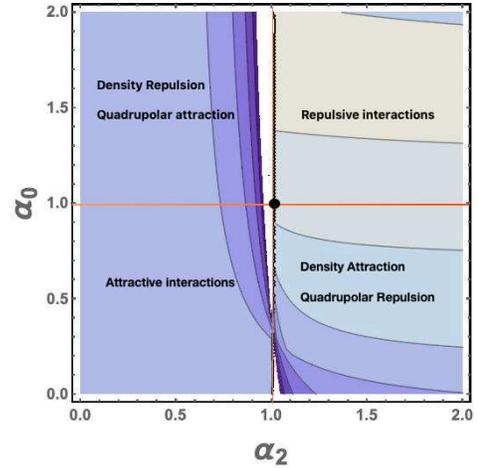}
\end{center}
\caption{Dynamical phase diagram in the $\alpha_2-\alpha_0$ plane, with $\alpha_0=1+F_0$ and $\alpha_2=1+F_2$. The central bold point $\alpha_0=\alpha_2=1$ correspond to the Fermi Gas.  
$\alpha_{0,2}>1$ ($\alpha_{0,2}<1$) represent repulsive  (attractive) interactions. The white region is the 
one where two stable collective modes are present. In the borders of this region there is sharp change in the collective modes structure, 
signaling a dynamical phase transition.}
\label{fig:pd}
\end{figure}

Repulsive isotropic interactions, $\alpha_0>1$, do not change the root structure. Indeed, the only effect is to increase the region with 
two stable modes. On the contrary, attractive interactions, $\alpha_0<1$, reduce this region that eventually disappears. 
These 
stable modes reappear for strong enough attractive isotropic interactions. This feature is depicted in Fig.~(\ref{fig:pd}). Here,  we show a 
dynamical phase diagram  in the  $\alpha_2-\alpha_0$ plane. The white area represent the region where two real roots separated from the 
particle-hole continuum are present. 
The border lines of these regions indicate a sudden change in the dynamical properties of the system signaling a dynamical phase transition.  
Indeed, the transition line in the attractive region  has a double pole in the Green functions, producing linear running out modes in real time. 

\subsection{Antisymmetric modes}
\begin{figure}[hbt]
\begin{center}
\subfigure[]
{\label{fig:rootsAnti-a}
\includegraphics[width=0.42\textwidth]{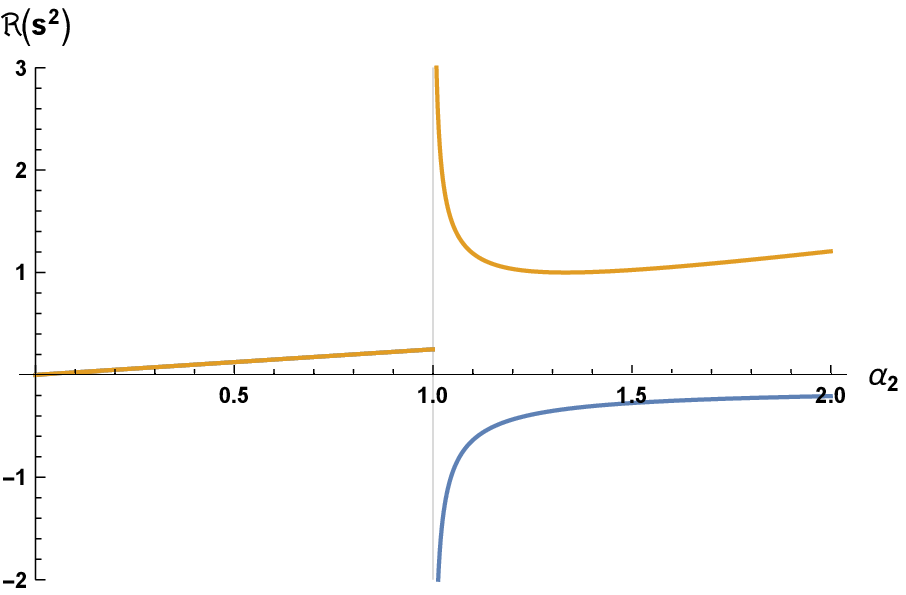}}
\subfigure[]
{\label{fig:rootsAnti-b}
\includegraphics[width=0.42\textwidth]{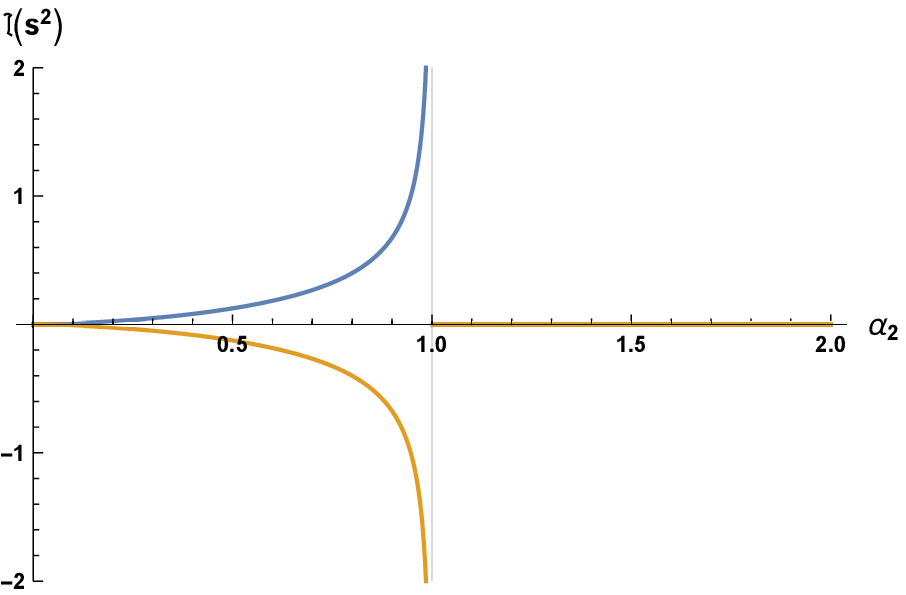}}
\end{center}
\caption{Antisymmetric collective modes. Solutions of the equation $D^-(s)=0$ with $D^-(s)$ given by Eq.~(\ref{eq:D-}). 
In panel~\ref{fig:rootsAnti-a} we plot $\Re[s^2(\alpha_2)]$ while in panel~\ref{fig:rootsAnti-b} we depict $\Im[s^2(\alpha_2)]$. 
In both panels we have fixed $\alpha_0=1$. }
\label{fig:rootsAnti}
\end{figure}

 In order to look for antisymmetric modes, we need to solve 
$
D^-(s)=0$, where  $D^-(s)$ is given by Eq.~(\ref{eq:D-}). 
Note that this expression  is $\alpha_0$ independent. We find the solutions
 \begin{equation}
s^2=\frac{\alpha_2}{4} \left\{
\begin{array}{lcl}
1\pm \sqrt{\frac{\alpha_2}{\alpha_2-1}} &  \mbox{for} & \alpha_2>1 \\
1\pm i\sqrt{\frac{\alpha_2}{1-\alpha_2}} &  \mbox{for} & \alpha_2<1
\end{array}
\right.
\end{equation}
which are depicted  in Fig.~(\ref{fig:rootsAnti}).
For repulsive quadrupolar interaction ($\alpha_2>1$), we have an overdamped and a stable mode. These  modes are not possible without higher 
angular momentum interactions. Note that both roots diverges in the limit $\alpha_2\to 1$. 
In the attractive region, there is a damped mode that runs all the way to the Pomeranchuk instability. 
In next section, we study in detail this region.

\subsection{Pomeranchuk regime}
The model described so far is reliable provided $\alpha_0>0$ and $\alpha_2>0$. Exactly at the values $\alpha_0=0$ or $\alpha_2=0$, 
the theory is unstable and the harmonic approximation is no longer valid.  These instabilities are known as Pomeranchuk instabilities. 
The $\alpha_2=0$ Pomeranchuk instability is associated with the quantum critical isotropic-nematic  phase transition~\cite{Lawler2006,OgKiFr2001}.  
Therefore, in order to understand this transition it is useful to know the structure of  collective modes near $\alpha_2\gtrsim 0$. 
 In our formalism, the instability is manifested by the appearance of a mode $s^2=0$ at $\alpha_2=0$. However, this is because we have chosen 
 a model with  highly localized interactions.  In fact, in Eq.~(\ref{eq:Sint}), we have considered 
 $F_{S,T}({\bf x}-{\bf x'})\equiv  F_{S,T}\delta^2({\bf x}-{\bf x'})$. It leads to constant parameters $\alpha_0,\alpha_2$.
 This model is not adequate near a Pomeranchuk instability and we need to consider finite ranged interactions. 
In this way, the parameters of the model are momentum dependent, {\em i.e.}, $\alpha_0\equiv \alpha_0(q)$ and $\alpha_2\equiv \alpha_2(q)$.
For short range interactions, it is possible to expand these couplings in Taylor series,  
\begin{align}
\alpha_0(q)&=\alpha_0+\kappa_0 q^2 \\
\alpha_2(q)&=\alpha_0+\kappa_2 q^2 ,
\end{align} 
 where $\kappa^2_{0}$ and $\kappa^2_{2}$ are  typical short distance ranges for density and quadrupolar interactions, respectively. 
 Deep in the Fermi liquid phase, these momentum dependent functions are irrelevant and it is sufficient to consider the zero momentum  
 constant Landau parameters. However, near Pomeranchuk instabilities, the momentum dependent terms are important. 
Expanding the symmetric as well as the antisymmetric collective modes  in powers of $\alpha_2(q)\gtrsim 0$ and setting the Pomeranchuk condition
  $\alpha_2=0$,   we find, 
\begin{align}
s^2& \sim \;\;  \frac{\alpha_0}{2}\;\;\;\;\;\;\;\;\;\to  \omega\sim \sqrt{\frac{\alpha_0}{2}} v_F q     \mbox{~~~~~~symmetric}  
\label{eq:linear}  \\
s^2& \sim    -\frac{\alpha_2^2(q)}{4}   \;\;\;\to  \omega\sim  i \frac{1}{2}\kappa_2  v_F q^3 \mbox{~~~~~ symmetric} 
\label{eq:cubic}  \\
s^2& \sim   \;\; \;  \frac{\alpha_2(q)}{4}     \;\;\;\to  \omega\sim  \frac{1}{2}\sqrt{\kappa_2}  v_F q^2 \mbox{~~~~antisymmetric}
\label{eq:quadratic}
\end{align} 
 We see that we have a stable linear mode, Eq.~(\ref{eq:linear}), and an overdamped cubic mode, Eq.~(\ref{eq:cubic}),  associated with the 
 symmetric collective modes. In addition, we have a stable quadratic mode, Eq.~(\ref{eq:quadratic}), associated with  antisymmetric collective modes. 
The overdamped cubic mode was identified as the most relevant to describe the isotropic-nematic phase transition, having its counterpart 
on the broken symmetry phase as a Goldstone mode~\cite{Lawler2006}.

\section{Spectral functions}
\label{sec:SpectralFuncions}
As already mentioned in the previous section, the Green function for the symmetric sector diverges as $s\to 0$. The reason is the existence of a 
zero eigenvalue of the matrix ${\bf M}^+$ in Eq.~(\ref{eq:I-M}). The existence of this zero mode does not depend on interactions, it is present 
even in the Fermi gas case, $\alpha_0=\alpha_2=1$. To understand the physical meaning of this mode,  it is convenient to turn back to the original 
equation, Eq.~(\ref{eq:Boltzman}). For static deformations, in the free case, we have 
$ \left({\bf v}_S\cdot {\bf q} \right)\delta n_S({\bf q},t)=0$. If $\left({\bf v}_S\cdot {\bf q} \right)\neq 0$ then, 
the only static solution is $\delta n_S({\bf q})=0$ for all $S$. However, if $\left({\bf v}_S\cdot {\bf q} \right)=0$, then there is a nontrivial 
solution $\delta n_S({\bf q})=\mbox{constant}$, with ${\bf q}\perp {\bf v}_S$. In this way, constant displacements perpendicular to Fermi 
velocity in each patch are solutions of the equation of motion. However, tangential displacements do not deform the Fermi surface, 
they are just reparametrizations of the same surface. 
It is instructive to understand this result in the angular momentum basis. 
From Eq.~(\ref{eq:1order}), it is simple to see that in the free case, there are nontrivial static solutions of the form
$m_{\ell-1}=-m_{\ell+1}$. This means, for instance, that $m_0=-m_{\pm 2}= m_{\pm 4}=-m_{\pm 6},\ldots$  are nontrivial solutions of the equation 
${\bf M^+ m^+}=0$.  Thus, the vector ${\bf m}^+=m_0(1,-1,1,-1,\ldots)$ is an eigenvector of ${\bf M}$ with zero eigenvalue. 
To understand the meaning of  this  eigenvector, we can replace it in Eq.~(\ref{eq:deltans-Fourier}).
We  find, 
\begin{align}
\delta n_S({\bf q})=m_0\sum_{\ell=-\infty}^{\infty} (-1)^\ell e^{i2\ell \theta_S}=m_0 \delta\left(\theta_S-\frac{\pi}{2}\right).
\end{align}
Thus, the zero mode  represent particle-hole excitations  with momentum ${\bf q}$ perpendicular to the Fermi velocity in each patch.   
It is simple to solve this problem in the angular momentum basis. The key observation is that, since the eigenvectors of the zero mode 
do not represent actual physical excitations,  the physical vector space is  the orthogonal subspace  to the zero mode eigenvector.  
Thus, we need to compute the  modified Green functions by restricting the vector space. 
This process is practically simple to implement. We need to expand the Green function in partial fractions   
 and define the modified Green functions by simply subtracting the contribution from the zero mode. Then, the modified Green function takes 
 the general form  
 \begin{align}
 {\bf G}_R(s,q)&= \sum_j \frac{B_j(s,q)}{(s^2-s^2_j)} \ ,
 \end{align}
 where $s^2_j\neq 0$ are the nontrivial roots computed in the previous section and the weights $B_j(s,q)$ are simply computed by expanding 
 the function in partial fractions.
 
\begin{figure}[hbt]
\begin{center}
\subfigure[]
{\label{fig:SFa}
\includegraphics[width=0.35\textwidth]{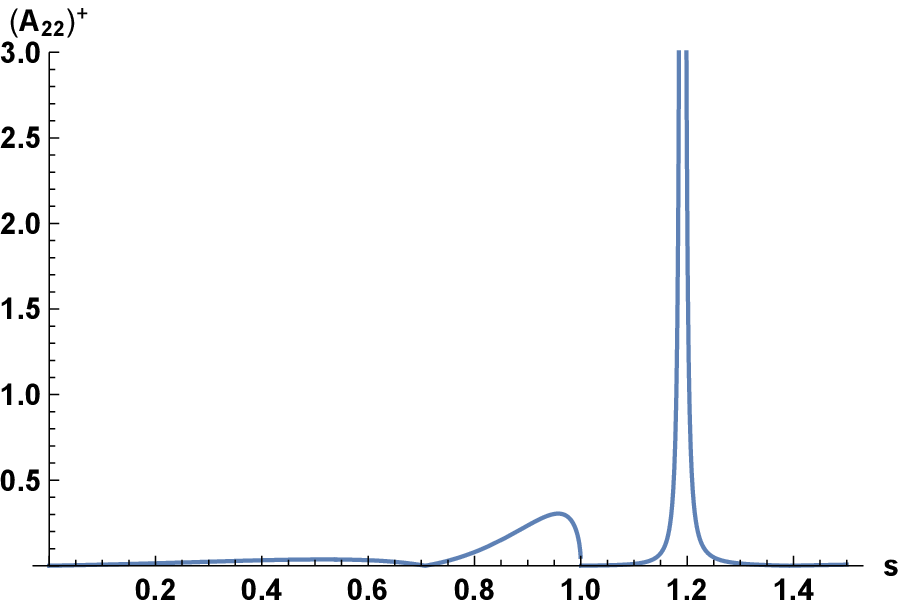}}
\subfigure[]
{\label{fig:SFb}
\includegraphics[width=0.35\textwidth]{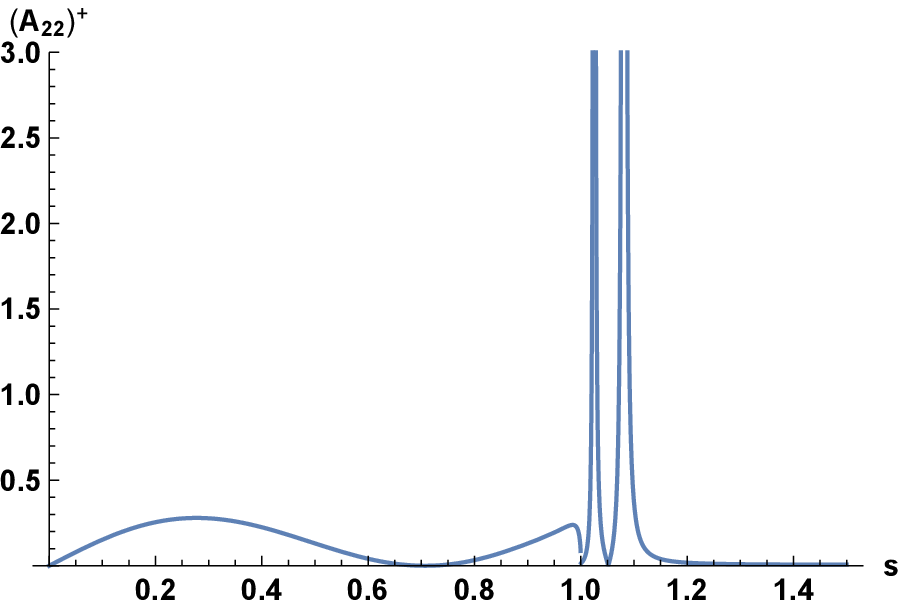}}
\subfigure[]
{\label{fig:SFc}
\includegraphics[width=0.35\textwidth]{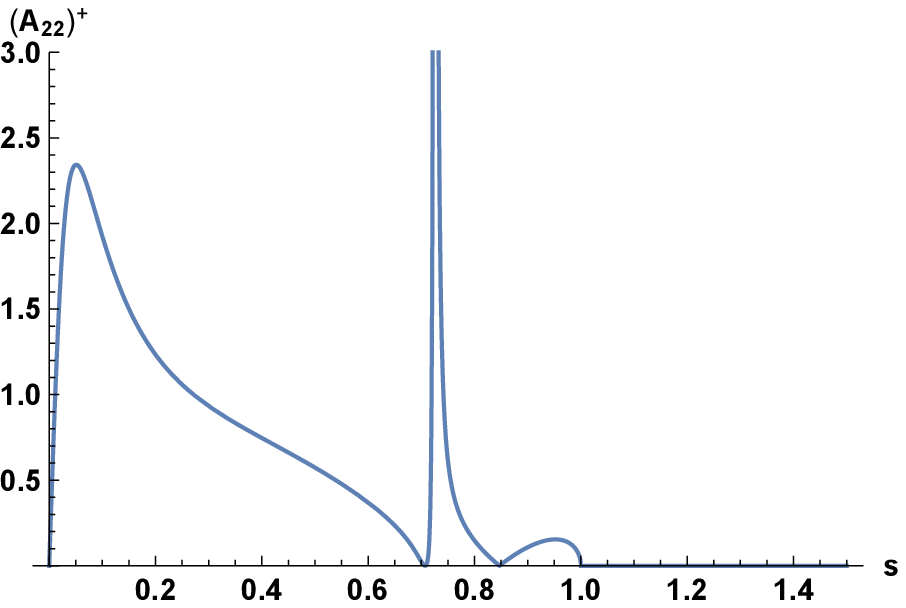}}
\end{center}
\caption{Normalized spectral function $A^+_{22}$ as a function of $s=\omega/v_Fq$, for $q$ fixed. We have fixed $\alpha_0=1$, in such a way that 
the roots coincide with those of Fig.~\ref{fig:roots-alpha2}. In \ref{fig:SFa} we have fixed $\alpha_2=2$, deep in the repulsive region. \ref{fig:SFb} is in the weak attractive region $\alpha_2=0.94$ (white region of Fig. \ref{fig:pd}), 
while in \ref{fig:SFc} we have fixed $\alpha_2=0.1$ near the Pomeranchuk instability.} 
\label{fig:SF}
\end{figure}

 We  computed the normalized spectral functions as 
\begin{equation}
{\bf A^{\pm}(\omega,q)}={\cal N}^{-1}\; \lim_{\epsilon\to 0^+} {\Im}[{\bf G_R}^{\pm}(\omega+i \epsilon, q)],
\end{equation}
where the normalization constant ${\cal N}^{-1}$ is chosen in such a way that 
$
\int_{-\infty}^{+\infty}{\bf A(\omega,q)} d\omega=1 
$
for fixed ${\vec q}$.
In Figure~\ref{fig:SF}, we show $A_{22}^+(s,{\bf q})$ for $\alpha_0=1$ and different values of the quadrupolar coupling  $\alpha_2$.
In panel~\ref{fig:SFa}  we fixed $\alpha_2=2$, deep in the repulsive region. In \ref{fig:SFb}, we show the spectral function for weak attraction 
$\alpha_2=0.94$, while the last panel \ref{fig:SFc} is in the strong repulsive regime, near the Pomeranchuk nematic instability,  $\alpha_2=0.1$.
In Fig.~\ref{fig:SFa}, we observe the quadrupolar  Landau zero sound as a sharp peak in the repulsive quadrupolar regime. Moreover, 
we observe a continuum particle-hole distribution for $s<1$.  This resonance moves to the left with decreasing $\alpha_2$ and reaches the free mode 
$s^2=1$ for $\alpha_2=1$. Figure~\ref{fig:SFb} shows an unexpected result. In the attractive region, there are two resonances. 
The resonance near $s\gtrsim 1$ is the continuation of the zero sound to the attractive region. However, there is a second mode, beginning at 
very high frequency that moves to lower frequency  for more attractive interactions. This behavior can be clearly observed in 
Fig.~\ref{fig:roots-alpha2}. Both resonances  meet each other at a critical $\alpha_{2c}\sim 0.93$, where there is a second order pole 
in the Green function. The location of this critical value depends on $\alpha_0$ and it is shown in Fig.~\ref{fig:pd}.  
For  smaller values of $\alpha_2$, the mode becomes damped and there is a huge transfer of spectral weight to the overdamped cubic mode, 
very near the Pomeranchuk instability.  

\begin{figure}[hbt]
\begin{center}
\subfigure[]
{\label{fig:SFAa}
\includegraphics[width=0.35\textwidth]{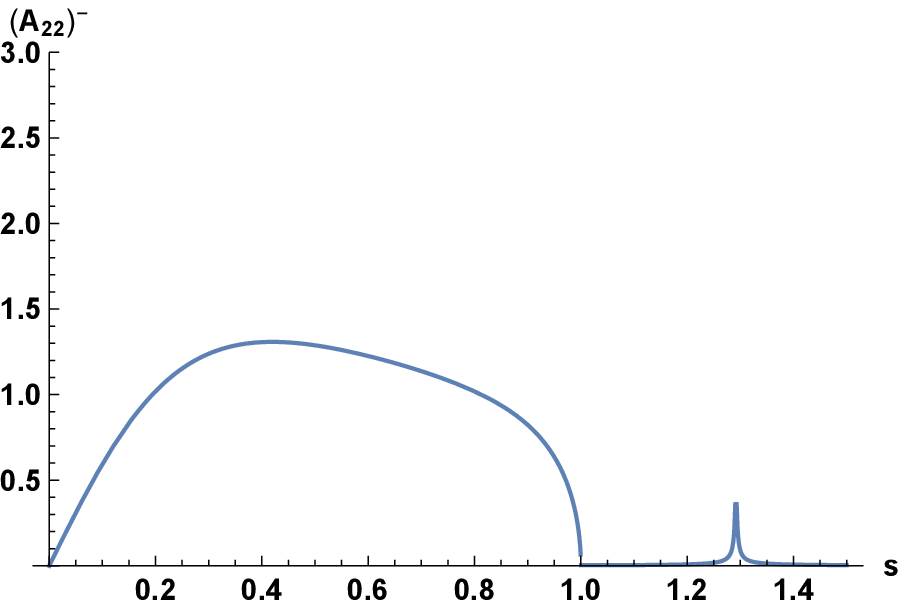}}
\subfigure[]
{\label{fig:SFAb}
\includegraphics[width=0.35\textwidth]{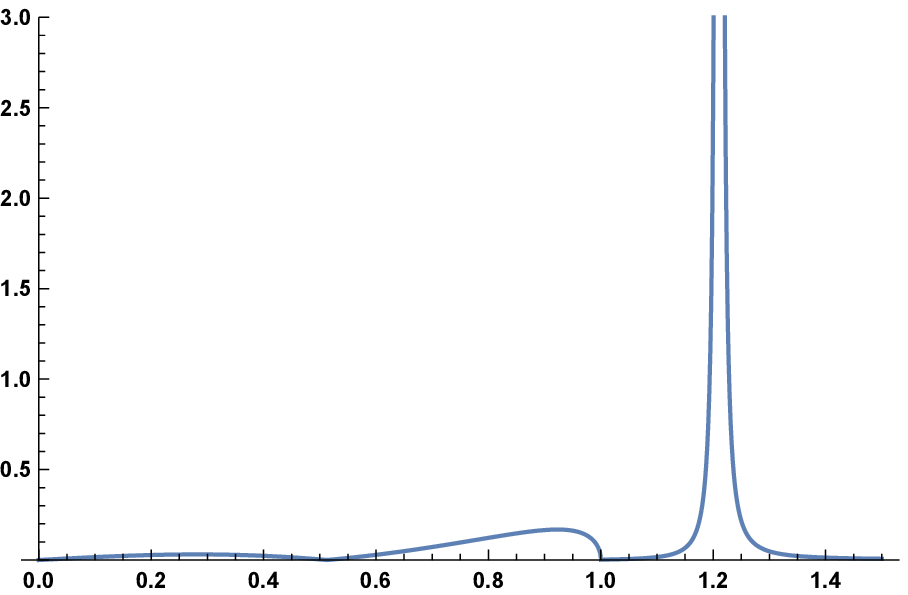}}
\subfigure[]
{\label{fig:SFAc}
\includegraphics[width=0.35\textwidth]{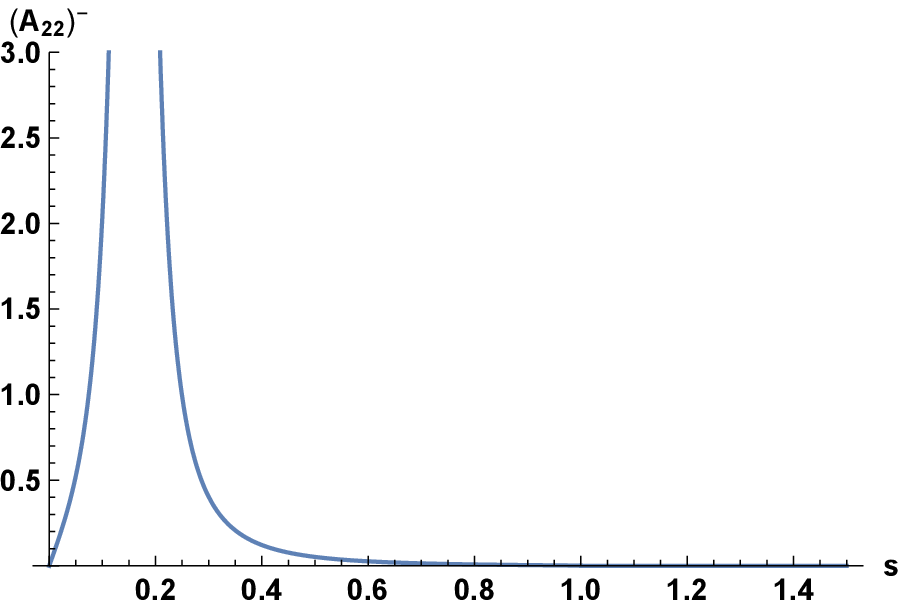}}
\end{center}
\caption{Normalized spectral function $A^-_{22}$ as a function of $s=\omega/v_Fq$, for $q$ fixed. The roots coincide with those of Fig.~\ref{fig:rootsAnti}. 
In \ref{fig:SFAa} we have fixed $\alpha_2=2$,, deep in the repulsive region.  Fig.  \ref{fig:SFAb} is computed for weak repulsion $\alpha_2=1.5$,  while in \ref{fig:SFc} we have fixed $\alpha_2=0.1$ near the Pomeranchuk instability.}  
\label{fig:SFA}
\end{figure}

We have also plotted in Fig.~\ref{fig:SFA}, the spectral function for the antisymmetric mode $A^-_{22}(s,q)$. 
Usually, in a Fermi liquid with isotropic interactions $F_0$, there are no sharp antisymmetric modes. However, they are possible when higher 
angular momentum interactions are considered. Antisymmetric modes were computed for Fermi liquis with interactions $F_0$ and $F_1$~\cite{Khoo2019} 
and at the nematic Pomeranchuk instability~\cite{Lawler2006}.  Here, we show that the same type of mode also appears with quadrupolar interactions 
in all the interaction regime.  However, for very strong repulsive interactions, the spectral weight of the share mode is very small, compared 
with the continuum particle-hole distribution, Fig.~\ref{fig:SFAa}. 
For small repulsion $\alpha_2=1.5$, this mode is sharply defined (see Fig.~\ref{fig:SFAb}) and is well separated from the particle-hole continuum 
for very weak repulsive interactions.  For attractive interactions, this mode  becomes damped, and very near the Pomeranchuk instability, this 
gets most of the spectral weight near $s\sim 0$, producing a well defined mode with a quadratic dispersion relation.

\section{Summary and discussions}
\label{Sec:Discussion}

We have presented a detailed study of collective modes of a Fermi liquid model with density and quadrupolar interactions. We have analyzed a 
wide range of interactions, from a repulsive regime to a strong attractive region in both interaction channels. We have used a  bosonization 
approach to Fermi liquids  to deduce a semiclassical evolution equation for Fermi surface fluctuations parametrized by an infinite  set of 
angular momentum modes.  
In this way, the system is mapped into a one-dimensional chain of angular momentum modes with first near neighbors interactions. 
Due to parity symmetry, longitudinal (symmetric) and transverse (antisymmetric) modes are decoupled. Moreover, odd and even angular momentum 
excitations are also decoupled.  
By means of a  recurrent decimation procedure, we have computed the relevant Green functions.  Exploring the  analytic properties,  
we were able to study collective modes and spectral functions for different values of the interaction 
parameters $\alpha_0=1+F_0$ and  $\alpha_2=1+F_2$.

In the repulsive region ($\alpha_0>1$, $\alpha_2>1$), we have found the expected Landau zero sound excitations,  modified 
by the presence of  quadrupolar interactions.  For antisymmetric (transverse) modes, we have also find a stable mode, 
compatible with a ``share sound'' recently reported in a  similar related system~\cite{Khoo2019}. However, we found that 
the spectral weight of this mode is highly suppressed for strong repulsion, having a clear signature above the particle-hole continuum 
for very weak repulsion, very near the Fermi gas region. It is important to note, that antisymmetric modes are not possible in Fermi 
liquids with only isotropic interactions $\alpha_0$.     

When attractive interactions are present ($\alpha_0<1$ or/and $\alpha_2<1$), the dynamical structure is more interesting. 
We found that weak quadrupolar attraction ($\alpha_0=1$, $\alpha_2\lesssim 1$) produces a sudden change in the symmetric modes dynamics. 
In addition with the continuation of the zero sound excitation, there is another stable mode at very high frequencies which moves to lower 
frequencies when the attraction growths. There is a critical value $\alpha_{2c}$, in which both stable modes meet together, 
melting in a single damped mode for stronger attraction ($\alpha_2<\alpha_{2c}$).
Exactly at $\alpha_{2c}$, there is a double pole in 
the Green function, producing a linear running out mode in real time.
For stronger attraction, besides the damped mode, 
an overdamped one gains spectral weight, being the dominant mode near the Pomeranchuk region $\alpha_2\gtrsim 0$. 
This behavior is sketched in Fig.~(\ref{fig:SF}).
This general collective modes structure remains the same in the presence of repulsive isotropic interactions $\alpha_0>1$. The effect of 
this coupling is to enlarge the intermediate region with two stable modes. However, attractive isotropic couplings $\alpha_0<1$ shrink 
this region which eventually disappears. Interestingly, an equivalent behavior is observed in a dual region $\alpha_0<1$ with $\alpha_2>1$, 
and remains unchanged  up to the 
Pomeranchuk region $\alpha_0\gtrsim 0$. We have illustrated this behavior in a dynamical phase diagram in Fig.~\ref{fig:pd}.
In this diagram, we show the lines where the structure of collective modes changes abruptly, possibly signaling a dynamical phase transition. 
The physics behind this transition can be traced back to the fact that the model behaves like a small quantum system, containing few degrees of freedom (an isotropic ``breathing mode"  and a quadrupolar "nematic" excitation),  in contact with a non-trivial  environment 
composed by an infinite set of  angular momentum excitations. The coupling  between the small effective system and the environment is governed 
by  $\sqrt{\alpha_2}$. The effective Hamiltonian  is non-hermitian,  signaling  a dissipative   behavior of the reduced system.  It is interesting to note that some small dissipative quantum systems present, depending on the environment,  dynamical phase transitions, characterized by crossing of energy levels, where some properties of the eigen-functions such as orthogonality and phase rigidity dramatically change\cite{Rotter2015}. 
The term ``dynamical phase transition'' is also used in another context to denote a lack of analyticity in  some time-dependent self-correlation functions. This point of view  is usually applied in the study of  closed system, where the time evolution is unitary. There should be  a deep relation between these two approaches. However, to the best of our knowledge, this issue  is not completely understood so far~\cite{Heyl2018}.

Summarizing, the main novel features we present in this paper are, in one hand, the observation of a sudden change in the collective 
mode structure which appears when the system mixes attractive and repulsive isotropic and quadrupolar  interactions. On the other hand, 
we show the existence of a sharp antisymmetric mode, very well defined for weak quadrupolar repulsion. 

The discotinuos dynamical behavior described in the paper will have an impact  on the out-of-equilibirum Fermi surface evolution in  quantum quenches set ups.\cite{Iucci2014}.
Although we have presented a simple phenomenological model, 
we believe that our results could also help to understand non-equilibrium responses of different systems that are being explored 
experimentally, especially those in which quadrupolar interactions seem to be relevant. 
Iron based superconductors are promising compounds~\cite{Fernandes2014,Li2017}, which have clear evidences of electronic nematic phases. 
Moreover, quantum Hall systems at partially filled Landau levels also display spontaneous breakdown of rotational 
symmetry~\cite{BaFrKiOg2002, FrKi1999,Wexler2001,Feldman2016}, signaling the relevance of the quadrupolar interaction. 
Since much of the  interesting features of our model  have signatures for weak  interactions,  we expect that ultracold Fermionic atoms~\cite{Chin2010,Schunck2005} could also be  an arena to look for this phenomenology, due to  the fact that  weak repulsive as well as attractive couplings can be manipulated with great precision in these systems.

\acknowledgments
We would like to acknowledge Eduardo Fradkin, Anibal Iucci and Zochil Gonz\'alez Arenas for  useful comments.
The Brazilian agencies, {\em Funda\c c\~ao de Amparo \`a Pesquisa do Rio
de Janeiro} (FAPERJ), {\em Conselho Nacional de Desenvolvimento Cient\'\i
fico e Tecnol\'ogico} (CNPq) and {\em Coordena\c c\~ao  de Aperfei\c coamento de Pessoal de N\'\i vel Superior}  (CAPES) - Finance Code 001,  are acknowledged  for partial financial support.
RA was partially supported by a MSc Fellowship from CAPES. 

\appendix
\section{Decimation procedure}
\label{App:Decimation}
In order to  compute the Green functions, Eqs.~(\ref{eq:M+}) and~(\ref{eq:M-}), we need to invert  an infinite range matrix. 
The tridiagonal form of these matrices greatly simplifies the calculation. Indeed, the mathematical model described by these matrices are completely 
analog to a linear chain of harmonic oscillators. The angular momentum in our problem is playing the role of the  ``lattice site'' in a 
tight-binding atomic chain.   
The decimation procedure is an effective algorithm to compute Green functions, specially in a one-dimensional lattice.  
The method can be summarized in the following way. The idea is to compute the inverse of a truncated matrix to order $n+1$, as a function 
of the inverse of the truncated matrix  of  order $n$. Provided this is  possible, the exact Green function is computed by means of a 
recurrence relation in the limit of $n\to\infty$. 

Consider, for instance, the matrix of Eq.~(\ref{eq:M+}). First, we truncate the matrix to order  $n=2$. Then, we need to compute
\begin{equation}
\left(
\begin{array}{cc}
s^2-\frac{\alpha_0}{2} & \frac{\sqrt{\alpha_0\alpha_2}}{2}    \\
 \frac{\sqrt{\alpha_0\alpha_2}}{4}    &   s^2-\frac{\alpha_2}{2}\\
\end{array}
\right)
\label{ap:N+2}
\left(
\begin{array}{cc}
    G_{00}^{+}(\omega, {\bf q}) & G_{02}^{+}(\omega, {\bf q})  \\
    G_{20}^{+}(\omega, {\bf q}) & G_{22}^{+}(\omega, {\bf q})
\end{array}
\right)
= {\bf I}.
\end{equation}
Let us focus in one component of the matrix, say $G_{00}^{+}$. 
Inverting the $2\times 2$ matrix, we easily find,
\begin{equation}
    G_{00}^{+ (2)} = \frac{1}{\displaystyle s^{2} - \frac{\alpha_{0}}{2} - 
    \frac{\alpha_{0}\alpha_{2}}{8}\frac{1}{ \displaystyle s^{2} - \frac{\alpha_2}{2}} }
\end{equation}
where the superscript $(2)$ means that the matrix was truncated to order 2.
Repeating the same procedure for truncations to order $n=3$ and $n=4$ we get, after direct calculation, 
\begin{align}
&G_{00}^{+(3)} = \frac{1}{\displaystyle s^{2} - \frac{\alpha_{0}}{2} - \frac{\alpha_{0}\alpha_{2}}{8}\frac{1}{ \displaystyle s^{2} - \frac{\alpha_{2}}{2} - \frac{\alpha_2}{16}\frac{1}{\displaystyle s^{2} - \frac{1}{2}}}} 
\label{ap:G3}\\
&G_{00}^{+ (4)} = \nonumber \\
&\frac{1}{\displaystyle
 s^{2} - \frac{\alpha_{0}}{2} - \frac{\alpha_{0}\alpha_{2}}{8}\frac{1}{\displaystyle s^{2} - \frac{\alpha_{2}}{2} - \frac{\alpha_2}{16}
\frac{1}{\displaystyle s^{2} - \frac{1}{2} - \frac{1}{16}
\frac{1}{\displaystyle s^{2} - \frac{1}{2}}} }}.
\label{ap:G4}
\end{align}
By carefully analyzing the recursive structure of $G_{00}^{+ (2)}$, $G_{00}^{+ (3)}$, $G_{00}^{+ (4)},\ldots$, we can 
write the $G^{+(n)}_{00}$ in the following way,  
\begin{align}
    G_{00}^{+ (n)} &= \frac{1}{\displaystyle s^2 - \frac{\alpha_{0}}{2} - \frac{\alpha_{0}\alpha_{2}}{8}\frac{1}{\displaystyle s^{2} - \frac{\alpha_{2}}{2} - \alpha_{2} \Pi^{(n)}(s)} } 
\end{align}
where  $\Pi^{(n)}(s)$ is given by the recurrence relation
\begin{equation}
\Pi^{(n+1)}(s) = \frac{1}{16}\frac{1}{\displaystyle s^{2} - \frac{1}{2} - \Pi^{(n)}(s)}
\label{ap:Pin}
\end{equation}
with the initial condition $\Pi^{(2)}(s)=0$.

The other components of the Green function,  $G_{02}^{+}$, $G_{20}^{+}$ and $G_{22}^\pm$ can be computed in the same way.  
We obtain for the symmetric modes   
\begin{equation}
{\bf \bar G}^{+(n)}(s)= \frac{1}{D^{+ (n)}(s)}\; {\bf N}^{+(n)}(s),
\label{ap:G+}
\end{equation}
where the numerator
\begin{equation}
{\bf N}^{+(n)}=\left(
\begin{array}{cccc}
s^2-\alpha_2(\frac{1}{2}+\Pi^{(n)}(s)) & & &   \frac{\sqrt{\alpha_0\alpha_2}}{2}    \\
 & & & \\
 \frac{\sqrt{\alpha_0\alpha_2}}{4}    &  &   & s^2-\frac{\alpha_0}{2}\\
\end{array}
\right)
\label{ap:N+}
\end{equation}
and the denominator
\begin{equation}
D^{+(n)}=\left[ s^{2} - \alpha_{2}\left( \frac{1}{2} + \Pi^{(n)}(s) \right) \right]\left(s^{2} - \frac{\alpha_{0}}{2}\right) - 
\frac{\alpha_{2}\alpha_{0}}{8}.
\label{ap:D+}
\end{equation}

Similarly, we obtain for the antisymmetric mode the Green function
\begin{equation}
G^{-(n)}_{22}(s)= \frac{N^{-(n)}(s)}{D^{-(n)}(s)},
\end{equation}
where 
\begin{equation}
N^{-(n)}=s^2-\frac{1}{2}-\Pi^{(n)}(s)
\label{ap:N-}
\end{equation}
and 
\begin{equation}
    D^{-(n)} = \left[ s^{2} - \frac{1}{2} - \Pi^{(n)}(s)  \right]\left(s^{2} - \frac{\alpha_{2}}{2}\right) - \frac{\alpha_{2}}{16} .
\label{ap:D-}
\end{equation}
Thus, we have inverted  the matrices of Eqs.~(\ref{eq:M+}) and~(\ref{eq:M-}) by truncating them at any finite order $n$. The explicit 
expression for these matrices are given in terms of a recursion relation for the function $\Pi^{(n)}$, given by Eq.~(\ref{ap:Pin}). 
To exactly compute the Green functions, we need to take the $n\to\infty$ limit. In that limit, $\Pi^{(n+1)}=\Pi^{(n)}$.  Defining 
\begin{equation}
\Pi(s)=\lim_{n\to\infty} \Pi^{(n)}(s), 
\end{equation}
 we find, from Eq.~(\ref{ap:Pin}),
\begin{equation}
\Pi(s) = \frac{1}{16}\frac{1}{s^{2} - \frac{1}{2} - \Pi(s)}
\label{ap:Pieq}
\end{equation}
which can be solved to get
\begin{equation}
    \Pi(s) = \left\{
\begin{array}{ccc}
\frac{1}{2}\left\{ s^{2} - \frac{1}{2} \pm |s|\sqrt{s^{2} - 1} \right\}  & \mbox{for} & s>1 \\
\frac{1}{2}\left\{ s^{2} - \frac{1}{2} \pm i |s|\sqrt{ 1-s^{2}} \right\}  & \mbox{for} & s<1 
\end{array}
\right.
\label{ap:Pi}
\end{equation}
Replacing $\Pi^{(n)}(s)$ by $\Pi(s)$ in  Eqs.~(\ref{ap:N+}), (\ref{ap:D+}), (\ref{ap:N-}) and~(\ref{ap:D-}), we find the exact Green 
functions presented in Eqs.~(\ref{eq:G+}) to~(\ref{eq:Pi}). 
 

%
\end{document}